\documentstyle[12pt,epsf,epsfig,wrapfig,cite]{article}
 \newcommand{\ddae}{$\vec{d}+d\to\!^4He+\eta~$}

\begin{document}
\begin{center}
{\bfseries MEASUREMENT OF THE VECTOR AND TENSOR POLARISATION\\ OF
 PROTON AND DEUTERON BEAMS\\}
\vskip 5mm
The GEM Collaboration\\
M.~Abdel-Bary$^{a}$, A.~Budzanowski$^{c}$, A.~Chatterjee$^{g}$, J.~Ernst$^{f}$, R. Gebel$^{a}$, P.~Hawranek$^{b}$, R.~Jahn$^{f}$, V.~Jha$^{g}$, K.~Kilian$^{a}$, S.~Kliczewski$^{c}$, Da.~Kirillov$^{a}$, Di.~Kirillov$^{k}$, D.~Kolev$^{e}$, M.~Kravcikova$^{j}$, T.~Kutsarova$^{d}$, \underline{M.~Lesiak$^{a,b}$}$^{\dag}$, J.~Lieb$^{h}$, H.~Machner$^{a,n}$, A.~Magiera$^{b}$, R.~Maier$^{a}$, G.~Martinska$^{i}$, S.~Nedev$^{l}$, N.~Piskunov$^{k}$, D.~Prasuhn$^{a}$, D.~Proti\'c$^{a}$, P.~von Rossen$^{a}$, B.~J.~Roy$^{g}$, I.~Sitnik$^{k}$, R.~Siudak$^{c,f}$, R.~Tsenov$^{e}$, M.~Ulicny$^{i}$, J.~Urban$^{i}$, G.~Vankova$^{a,e}$, C.~Wilkin$^{m}$ \\
\vskip 5mm
{\small
$^a${\it Institut f\"{u}r Kernphysik, Forschungszentrum J\"{u}lich, 52425
J\"{u}lich, Germany}\\
$^b${\it Institute of Physics, Jagiellonian University, Krak\'ow, Poland}\\
$^c${\it Institute of Nuclear Physics, Polish Academy of Sciences,
Krak\'ow, Poland}\\
$^d${\it Institute of Nuclear Physics and Nuclear Energy, Sofia,
Bulgaria}\\
$^e${\it Physics Faculty, University of Sofia, Sofia, Bulgaria}\\
$^f${\it Helmholtz-Institut f\"{u}r Strahlen- und Kernphysik der Universit\"{a}t
Bonn, Bonn, Germany}\\
$^g${\it Nuclear Physics Division, BARC, Mumbai, India}\\
$^h${\it Physics Department, George Mason University, Fairfax, Virginia,
USA}\\
$^i${\it P.~J.~Safarik University, Kosice, Slovakia}\\
$^j${\it Technical University, Kosice, Kosice, Slovakia}\\
$^k${\it Laboratory for High Energies, JINR Dubna, Russia}\\
$^l${\it University of Chemical Technology and Metallurgy, Sofia,
Bulgaria}\\
$^m${\it Department of Physics \& Astronomy, UCL, London, U.K.}\\
$^n${\it Fachbereich Physik, University Duisburg-Essen, Germany}\\
\vskip 5mm
$\dag$ {\it E-mail: m.lesiak@fz-juelich.de} }
\end{center}
\vskip 5mm

\begin{abstract}
Measurement of the \ddae reaction using vector and tensor polarised
beam has been performed at COSY using Big Karl magnetic
spectrograph. The beam polarisation necessary for
obtaining the vector and tensor analysing power for this reaction was measured. The method and the results of the tensor polarisation
measurement of the deuteron beam are presented.
\end{abstract}
\vskip 8mm

\section{Introduction}
There is a great interest in $\eta$-physics in recent years
\cite{Bij02}. The completed data set for $p+d\to\!^3He+\eta$ reaction
\cite{betigeri01,faeldt02} in a wide beam energy range exist but the related
channels of $d+d\to\!^4He+\eta$, $n+\!^3He\to\!^4He+\eta$ and
$p+t\to\!^4He+\eta$ are poorly measured. Whereas for the $p+t$ and
$n+\!^3He$ collision no data were taken at all due to the problems
with neutron beam and triton target, the existing data for the
$d+d\to\!^4He+\eta$ reaction are so far limited to the total and
differential cross section at small excess energies ($Q < 7.7$ MeV)
\cite{frascaria94,willis97,ola05}.

There are many theoretical models describing the $p+d\to\!^3He+\eta$
reaction. The two step model proposed by Kilian and Nann~\cite{Kil90} was
extended by F\"aldt and Wilkin~\cite{Wil96} to describe the
$d+d\to\!^4He+\eta$ reaction. However due to the lack of data the
question about the underlying reaction mechanism can not be answered yet.

The GEM collaboration performed an experiment investigating \ddae reaction using unpolarised as well as vector and tensor polarised beams at $Q = 17.5$ MeV. The measurement of the angular dependence of the differential cross section and the analyzing powers for the \ddae reaction allows to extract the partial wave amplitudes directly from the experimental data, not relying on any theoretical model. The energy dependence of absolute values of the partial wave amplitudes together with their phases obtained from the measured observables can be used to determine the reaction mechanism.

The vector polarisation of the beam was measured using a low energy
polarimeter placed in the beam line between the cyclotron and the COSY
accelerator. The tensor polarisation was measured at the external target
station using Big Karl magnetic spectrograph. The tensor polarisation $p_{zz}$
of the deuteron beam was extracted using deuteron-proton backward elastic scattering.

\section{Experiment and results}

\begin{figure}[!ht]
\begin{center}
\includegraphics[width=8cm]{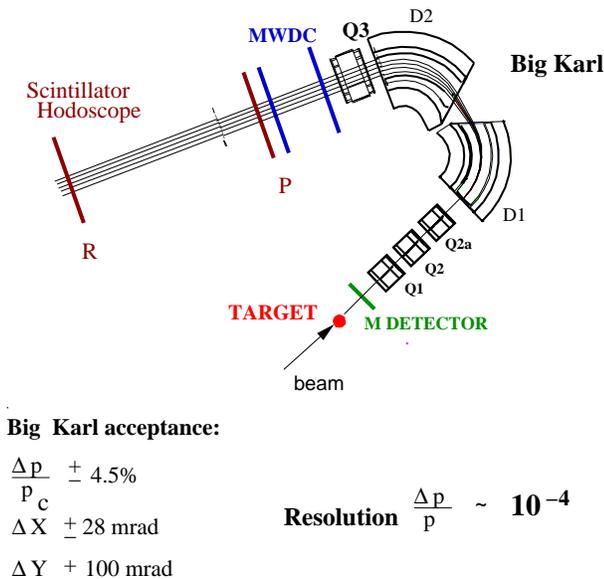}
\begin{footnotesize}
\caption{{\em Schematic layout of the detection system of Big Karl
used in the experiment. Q indicates magnetic quadrupoles lenses, D magnetic dipoles.}}
\label{bk}
\end{footnotesize}
\end{center}
\end{figure}

\begin{figure}[!ht]
\begin{center}
\includegraphics[width=8cm]{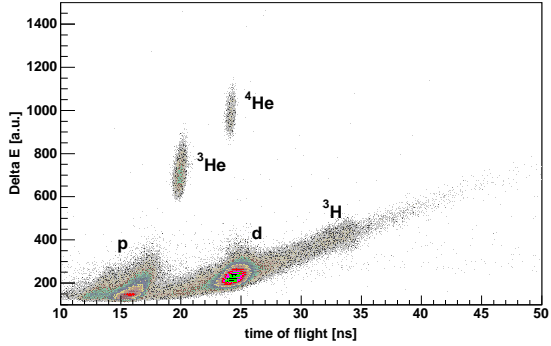}
\begin{footnotesize}
\caption{{\em Time of flight versus energy losses in scintillator layers for the $d(\vec d,X)$ reaction at 1.16 GeV.}}
\label{ap_tof}
\end{footnotesize}
\end{center}
\end{figure}

The experiment was performed at the COSY
\footnote{http://www.fz-juelich.de/ikp/de/} facility in J\"ulich using the Big
Karl magnetic spectrograph \cite{drochner98}. The detector system is
schematically shown in Fig. \ref{bk}. The Big Karl spectograph is equipped
with two sets of multi-wire drift chambers (MWDC) for position measurement and
two layers of scintillating hodoscopes for time of flight and energy losses
information, used for particle indentification. Very good particle separation
was achieved as it is seen in Fig. \ref{ap_tof}.

In the measurement of the $\vec{d}+ d \rightarrow \alpha + \eta$ reaction
deuteron beam of 1.16 GeV kinetic energy was used. Apart from the unpolarised beam, two different combinations of vector $p_{z}$ and tensor $p_{zz}$ polarisation were applied:
\begin{equation}\label{notation-1}
\small{p_{z} = -\frac{1}{3} \qquad p_{zz} = +1},
\end{equation}
\begin{equation}\label{notation-1}
\small{p_{z} = -\frac{1}{3} \qquad p_{zz} = -1}.
\end{equation}

The vector polarisation $p_{z}$ was measured with a low energy polarimeter placed in the injection beam line where the deuteron energy is about 76 MeV. Using carbon target, proton elastic scattering was measured with scintilatting detectors. The results for both polarisation combinations are: $p_{z} = -0.32 \pm 0.02$ and $p_{z} = -0.33 \pm 0.02$, respectively.

The tensor polarisation of the beam was measured using deuteron beam and liquid hydrogen target. The deuterons originating from $\vec d p$ backward elastic scattering were measured with Big Karl magnetic spectrograph.

The cross section for $\vec d p$ backward elastic scattering can be expressed as:
\begin{eqnarray}
\small{ \Bigg[ \frac{d\sigma}{d\Omega}(\theta=180^{o}) \Bigg]_{pol} = \Bigg[ \frac{d\sigma}{d\Omega}(\theta=180^{o}) \Bigg]_{unpol}\Bigg[ 1 - \frac{1}{2}t_{20}T_{20}(\theta=180^{o}) \Bigg]},
\label{dsdo_pzz}
\end{eqnarray}
where tensor polarisation can be written as:
\begin{eqnarray}
\small{p_{zz} = \sqrt2 t_{20}}.
\label{pzz}
\end{eqnarray}

\begin{figure}[!ht]
\begin{center}
\includegraphics[width=10cm]{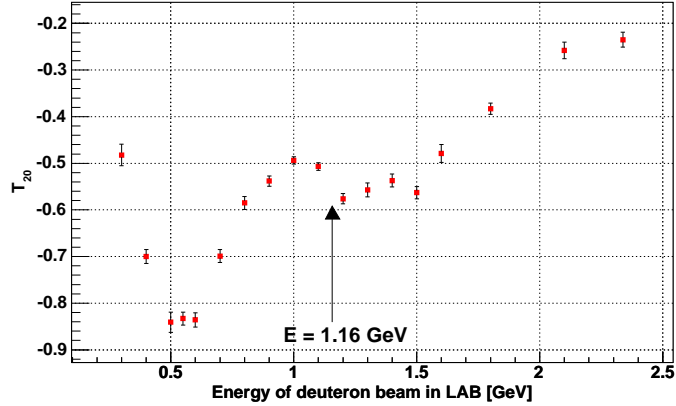}
\begin{footnotesize}
\caption{{\em Tensor analyzing power $T_{20}$ at $180^{o}$ measured by Ref. \cite{punj95}. The arrow indicates the beam energy of the present experiment.}}
\label{punjT20}
\end{footnotesize}
\end{center}
\end{figure}

\noindent
In order to determine $p_{zz}$, polarised and unpolarised cross sections as well as tensor analyzing power $T_{20}$ at $180^{o}$ have to be known.
 The tensor analyzing power $T_{20}$ at $180^{o}$ was already determined by the Saclay group \cite{arv83}. They measured protons from $\vec d p$ backward elastic scattering in a broad deuteron beam energy, covering our beam energy at 1.16 GeV. The data points were remeasured with smaller ucertainties by Punjabi et al. \cite{punj95} and these results are presented in Fig. \ref{punjT20}. The cross sections for different polarisation combinations were measured with Big Karl.

\begin{figure}[!ht]
\begin{center}
\includegraphics[width=8cm]{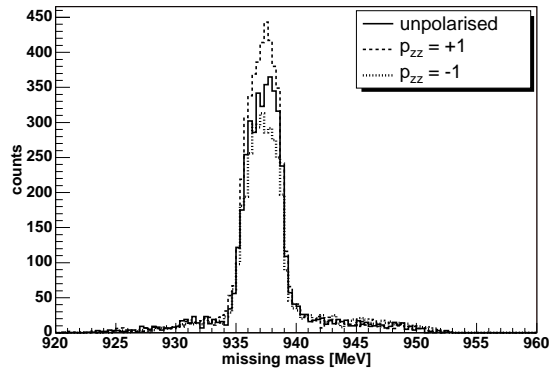}
\begin{footnotesize}
\caption{{\em Missing mass spectrum for $\vec d p \rightarrow pd$ reaction for unpolarised beam (solid line) and two different tensor polarisations of the beam (dashed and dotted lines).}}
\label{mm013}
\end{footnotesize}
\end{center}
\end{figure}

The outgoing deuterons were identified using the specific energy losses and
time of flight. In order to eliminate background, the deuteron  missing
mass spectra were analysed. An example of such a spectrum is shown in
Fig. \ref{mm013} for $\vec d +p \rightarrow d + X$ process for unpolarised
beam and for both polarisations of the beam. From these information the
differential cross sections for unpolarised beam and both polarisations were obtained. Using the relation (\ref{dsdo_pzz}) and (\ref{pzz}), tensor
polarisations were determined: $p_{zz} = +0.81 \pm 0.16 \pm 0.01$ and $p_{zz}
= -0.60 \pm 0.11 \pm 0.01$, where the first error is statistical and the
second is a systematical one. The results of vector and tensor polarisation
measurement are summarized in Table \ref{pzz_result}.

\begin{table}[!ht]
\begin{center}
\begin{tabular}{|c|c|}
                    \hline \hline
                    $p_{z}$ & $p_{zz}$ \\
  nominal \hspace{1.5cm} measured & nominal \hspace{1.5cm} measured\\
                    \hline \hline
-1/3 \hspace{1.5cm} -0.33 $\pm$ 0.02 & -1 \hspace{1.5cm} -0.60 $\pm$ 0.11 $\pm$ 0.01 \\
\hline
-1/3 \hspace{1.5cm} -0.32 $\pm$ 0.02 & +1 \hspace{1.5cm} +0.81 $\pm$ 0.14 $\pm$ 0.01 \\
\hline \hline
\end{tabular}
\caption{Results of the vector and tensor polarisations measurement.}
\label{pzz_result}
\end{center}
\end{table}

\vspace{13pt} \centerline{\bf{Acknowledgement}}
\vspace{13pt}

This work was supported by the Research Centre J\"ulich, the European Community-Research Infrastructure Activity under the FP6 "Structuring the European Research Area" programme (HadronPhysics, contract number RII3-CT-2004-506078) and Int. B\"uro BMBF (DLR) contract IND-01/022.

\end{document}